# SIMULATION DU COMPORTEMENT DYNAMIQUE DU VOILIER

## *DYNAMIC BEHAVIOUR SIMULATION OF SAILING BOAT*

### K. RONCIN

Laboratoire de Mécanique des Fluides UMR CNRS 6598
Division Hydrodynamique Navale,
École Centrale de Nantes, BP 92101, 44321 NANTES Cedex 3

**Résumé**

Un simulateur de deux voiliers en interaction est réalisé dans l'optique de fournir au sportif un outil d'analyse fiable. Nous décrivons ici les différents modèles utilisés. Un soin particulier a été apporté à l'évaluation des efforts hydrodynamiques stationnaires. La méthode des plans d'expériences a été mise en œuvre pour exploiter les essais en bassin des carènes réalisés au Laboratoire de Mécanique des Fluides de l'École Centrale de Nantes. Des résultats partiels du modèle obtenu sont présentés. Enfin, un exemple de simulation illustre les possibilités offertes.

**Summary**

A simulator of two sailing boats in interaction is being made in order to provide sportsmen with a reliable analytical tool. We presently describe different models we used. Stationary hydrodynamics efforts have been specially evaluated. Experiences plan method has been implemented to exploit the towing tank tests made in the Fluids Mechanic Laboratory of Ecole Centrale de Nantes. Here are set out partial results of the model we achieved. At last, a simulator example illustrates the possibilities provided.

# 1 INTRODUCTION

L'étude présentée est extraite d'un travail thèse en cours au Laboratoire de Mécanique des Fluides de l'École Centrale de Nantes. Son objectif est de fournir au sportif un outil d'analyse fiable pour l'optimisation de la conduite d'un voilier en match racing. Dans cet article, nous abordons les différentes étapes de la construction du simulateur. Le voilier évolue à la frontière de deux domaines fluides, l'air et l'eau. Les deux aspects fondamentaux en sont donc l'analyse des actions aérodynamiques sur les voiles et les œuvres mortes et l'analyse des actions hydrodynamiques sur la carène et les appendices (quille et safran). Bénéficiant de l'accès à un bassin d'essais des carènes, nous avons naturellement privilégié la modélisation des efforts hydrodynamiques. Un simulateur nécessite des données structurées. Cela nous a incité à aborder les essais en bassin en mettant en œuvre la technique des plans d'expériences, encore inexploitée dans ce domaine.

La plupart des études sur les voiliers sont orientées vers la conception. Il s'agit de déterminer l'influence des principaux paramètres de forme pour optimiser les performances. Notre optique est différente. Nous nous attachons à décrire le comportement d'un voilier donné pour optimiser son fonctionnement et aborder des problèmes simples de tactique. Dès lors, nous pouvons approfondir l'étude de l'influence des paramètres de position et d'attitudes, là où les études classiques se contentent de modèles simplifiés, (pour l'influence de la gîte par exemple), ou même éludent le problème comme pour l'influence de l'assiette.

Dans cet article nous décrivons notre démarche vis à vis de chaque composante du simulateur et les modèles que nous choisissons. Nous présentons comme résultats des éléments du modèle hydrodynamique issus des essais et enfin nous commentons un exemple de résultat obtenu avec le simulateur pour le deux voiliers en interaction.

**NOTATIONS ET REPERES**

$\vec{V}_B$ : vitesse du centre de gravité du voilier dans le référentiel absolu ($R_0$).

VMG (Velocity Made Good) : projection de $\vec{V}_B$ sur l'axe du vent réel.

$C_L$ : coefficient de portance.     $V_{WA}$ : vitesse du vent apparent.
$C_D$ : coefficient de traînée.     $V_{WT}$ : vitesse du vent réel.
$C_T$ : coefficient d'effort total.     $\beta_{WA}$ : angle du vent apparent.
$\rho_A$ masse volumique de l'air.     $\gamma$ : ratio traînée sur portance pour le gréement (voiles, mat, etc.).
$\delta$ : angle: de barre     Angles de cardan : $\phi$ gîte (roulis), $\theta$ assiette (tangage), et $\psi$ cap (lacet).

$R_0$ : repère galiléen lié à la terre, origine sur la surface libre au repos, l'axe $X_0$ dans la direction du nord géographique, $Z_0$ vertical ascendant.

$R_B$ : repère lié au bateau, origine au centre de gravité nominal, $X_B$ vers l'avant, $Y_B$ vers bâbord, $Z_B$ vers le haut.

La position du voilier, assimilé à un solide, est connue à tout instant par le vecteur position et par $\phi$, $\theta$, et $\psi$.

Le vecteur position : $\vec{P} = \overrightarrow{O_0 O_b} = \begin{pmatrix} x \\ y \\ z \end{pmatrix}$

L'attitude est : $\begin{pmatrix} \phi \\ \theta \\ \psi \end{pmatrix}$ et le vecteur rotation instantané dans ($R_B$) s'exprime par :

$\begin{pmatrix} p \\ q \\ r \end{pmatrix} = [\Sigma] \bullet \begin{pmatrix} \dot{\phi} \\ \dot{\theta} \\ \dot{\psi} \end{pmatrix}$ ; avec $[\Sigma] = \begin{bmatrix} 1 & 0 & -\sin\theta \\ 0 & \cos\phi & \cos\theta * \sin\phi \\ 0 & -\sin\phi & \cos\theta * \cos\phi \end{bmatrix}$

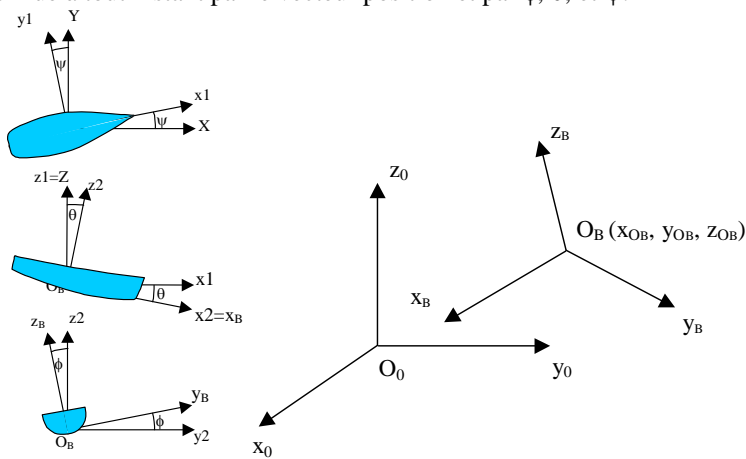

## 2   DEMARCHE POUR LA REALISATION DU SIMULATEUR

Le voilier est assimilé à un solide rigide. De façon schématique, le principe fondamental de la dynamique s'écrit dans le repère lié au bateau :

$$\left\{\begin{array}{l}\text{efforts}\\ \text{hydrodynamiques}\end{array}\right\}_{O_B,R_B} + \left\{\begin{array}{l}\text{efforts}\\ \text{aérodynamiques}\end{array}\right\}_{O_B,R_B} + \left\{\begin{array}{l}\text{efforts hydrostatiques}\\ \text{et de  pesanteur}\end{array}\right\}_{O_B,R_B} = \left\{\begin{array}{l}\text{Torseur}\\ \text{dynamique}\end{array}\right\}_{O_B,R_B}$$

Un module d'évaluation est construit pour chaque type d'effort. La méthode utilisée pour réaliser les modules est fonction des moyens et des modélisations physiques disponibles. Le degré de complexité de chaque modèle est également choisi pour que le temps de calcul reste compatible avec les différentes utilisations pratiques du simulateur.

Le travail de modélisation est réalisé en priorité avec les outils expérimentaux et numériques disponibles ou développés au Laboratoire de Mécanique des Fluides de l'École Centrale de Nantes.

L'option retenue a été de construire le simulateur de façon à pouvoir intégrer les différents modèles nécessaires à son fonctionnement au fur et à mesure de leur évolution, que les progrès soient obtenus par des études numériques ou expérimentales. L'environnement Matlab-Simulink a été adopté parce qu'il permet de dissocier dans le simulateur des entités indépendantes et interchangeables. Chaque entité est définie structurellement et visuellement par un objet. Elle est reliée aux autres entités par des relations d'entrées-sorties. Ce mode de programmation est bien adapté à l'objectif de modularité.

## 3   EFFORTS DE PESANTEUR ET EFFORTS HYDROSTATIQUES

Les efforts de pesanteur sont exprimés par leur résultante en $O_B$. La masse des équipiers et leurs positions sont introduites indépendamment pour permettre d'intervenir sur le réglage de l'attitude du bateau (gîte et assiette). Les effets dynamiques du déplacement des équipiers sur le bateau ne sont pas pris en compte pour l'instant

Les efforts hydrostatiques sont obtenus par un modèle en fonction des paramètres d'attitude. Dans la version actuelle le modèle est linéaire mais il n'y aura aucune difficulté à l'améliorer.

## 4   EFFORTS HYDRODYNAMIQUES

Les efforts hydrodynamiques sont la résultante de l'action des contraintes normales et tangentielles sur la partie de la coque en contact avec l'eau. Du fait des mouvements du bateau et de l'état de la mer l'écoulement réel est en général complexe. L'approche habituelle est de décomposer les efforts hydrodynamiques en trois types d'efforts dont les effets sont cumulés.

- Les efforts engendrés par le déplacement du bateau en translation à vitesse constante et attitude fixe. Ce fonctionnement correspond à la marche stabilisée du voilier sur eau plate.
- Les efforts et les mouvements du bateau provoqués par les vagues.
- Les efforts engendrés par les manœuvres du bateau, changement de cap, d'attitude et de vitesse. Nous les appellerons efforts de manœuvrabilité, ils comprennent une partie stationnaire, avec les efforts engendrés par les rotations, et une partie non-stationnaire fonction entre autres des accélérations.

Dans la réalité du fonctionnement du voilier, tous ces efforts hydrodynamiques coexistent et sont couplés à des degrés divers.

## 4.1 LES EFFORTS PROVOQUES PAR LES VAGUES

Les bateaux à voiles ont en général une longueur de l'ordre de la longueur d'onde des vagues significatives ou inférieure. Dans ces conditions, l'état de mer à une influence directe sur la façon dont les barreurs réalisent les manœuvres. L'idéal serait donc de connaître à chaque instant les efforts appliqués au bateau. Mais le problème est difficile et dans un premier temps nous nous intéresserons uniquement aux effets moyens. Il n'est alors pas utile de connaître les efforts et les mouvements dans le repère moyen à chaque instant. Il suffit de déterminer globalement la moyenne de l'augmentation de la résistance à l'avancement. (et la modification des efforts moyens sur les voiles). Cela demande une étude préalable pour relier ces grandeurs aux caractéristiques du bateau, à l'état de la mer, à la vitesse et au cap du bateau par rapport à la direction des vagues.

Cette étude n'étant pas encore réalisée pour le First-Class 8, les simulations présentées ne tiennent pas compte de l'état de la mer.

## 4.2 EFFORTS DE MANŒUVRABILITE

Les efforts engendrés par les manœuvres du voilier dépendent de façon couplée de la position et des angles d'attitude ainsi que de leurs dérivées premières et secondes. Les formulations simplifiées ne prennent en compte que les coefficients des termes d'ordre le plus bas dans le développement de ces efforts. Le principal problème vient des difficultés d'évaluation des coefficients et principalement de ceux qui concernent les vitesses de rotation. Leur évaluation par le calcul n'est pas encore opérationnelle et l'évaluation expérimentale fait appel à des procédures et à du matériel très lourds dont nous ne disposons pas. Pour l'instant nous prenons en compte ces effets sur les surfaces portantes en transportant le torseur cinématique au centre d'effort de ces surfaces. Pour la carène nous utilisons des coefficients extrapolés de valeurs obtenues par Masuyama dans [15] [16] pour un voilier de taille légèrement supérieure mais de conception voisine.

Les premiers tests avec le simulateur montrent que la valeur des coefficients des termes de vitesse de rotation de la carène a peu d'influence sur les résultats contrairement aux effets portants sur les appendices. En conséquence nous n'avons pas cherché à approfondir la détermination de ces coefficients, d'autant plus que nous pourrons prochainement utiliser le code Icare [2] qui permettra d'identifier tous les coefficients de manœuvrabilité sans réaliser d'expérimentation.

## 4.3 EFFORTS HYDRODYNAMIQUES STATIONNAIRES

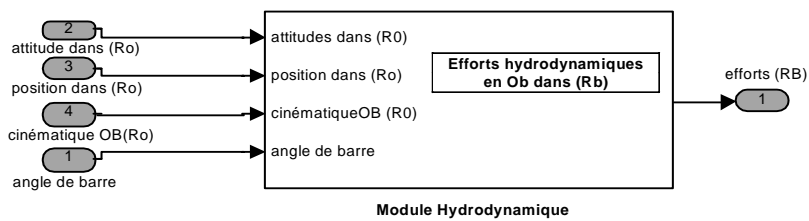

**Figure 1**

La qualité de l'évaluation des efforts hydrodynamiques conditionne la précision de la performance globale du voilier. Des outils numériques de calcul d'écoulement autour des carènes, en code « Fluide Parfait » (REVA [6], Aquaplus) et en code Navier-Stokes [2] sont développés et perfectionnés par un travail de recherche permanent. Nous avons cependant privilégié dans un premier temps la démarche purement expérimentale. Le Laboratoire dispose en effet de moyens d'essais lourds en hydrodynamique, notamment un bassin des carènes de 72 m dont on va prochainement doubler la longueur, un dynamomètre à 6 composantes et un orienteur spécifique pour l'étude des voiliers [23]

*4.4  UN POINT SUR LES METHODES EXPERIMENTALES*

De nombreuses techniques ont été mises en œuvre pour tester les voiliers en bassin des carènes. Elles vont de la technique en navigation libre (« free sailing technique ») dont le principe est d'appliquer au bateau une force censée reproduire l'effort vélique, à la technique captive, qui consiste à tracter le bateau à une vitesse donnée en bloquant tous ses degrés de liberté pour lui imposer une attitude fixe. Avec la première méthode on mesure, lorsque qu'elle est stabilisée, l'attitude prise par le bateau. Avec la deuxième on mesure le torseur des efforts appliqués au bateau.

Dans la pratique les techniques de mesure adoptées sont des variantes de ces deux méthodes extrêmes Elles sont obtenues, suivant le cas, en bridant ou en relâchant certains degrés de liberté.

La méthode en navigation libre a été introduite par Alan et al. en 1957 [1]. L'avantage recherché est d'obtenir des combinaisons réalistes des paramètres d'attitude et de diminuer ainsi le nombre d'essais nécessaire pour tester une carène. La technique la plus récente est développée au MARIN [8]. La maquette n'est pas tout à fait libre puisque l'angle de dérive est imposé. Le point de traction se déplace horizontalement grâce à des servo-moteurs pour éliminer la composante de moment de lacet. Le temps de stabilisation est de 15 à 20 secondes. Bien que séduisantes ces méthodes présentent quelques inconvénients. En premier lieu les attitudes obtenues dépendent de suppositions sur la force propulsive appliquée et en particulier sur le point d'application de cette force. Elles dépendent aussi de l'orientation du safran qui parfois est introduite comme paramètre indépendant. D'autre part, la méthode ne peut pas fournir de données pour des combinaisons de paramètres d'attitudes correspondant à un bateau en manœuvre. Enfin, l'état stable du système, qui parfois n'est pas atteint, met un certain temps à s'établir. D'un point de vue technique ces méthodes obligent à respecter la position verticale du centre de gravité sur la maquette et demandent des dispositifs expérimentaux sophistiqués.

Après dix ans d'expérience de ce type d'essais Murdey au NRC [18] conclut que les méthodes d'essais semi-captives sont en définitive plus productives à condition de les associer avec un VPP. Elles permettent de tester plusieurs états d'équilibre pour une carène.

Les méthodes semi-captives reviennent à fixer le bateau sur un dynamomètre à l'aide d'un dispositif qui permet de libérer un ou plusieurs degrés de liberté. Ces degrés de liberté correspondent évidemment aux mouvements qui présentent une raideur de sorte qu'un état d'équilibre puisse être atteint en cours de mesure. Il s'agit donc du pilonnement, du tangage et éventuellement du roulis.

A l'E.C.N. C. Talotte [23] a mis au point une technique laissant libre ces trois degrés de liberté et appliquant la force de traction à une hauteur proche de celle du point d'application des efforts véliques. Les instabilités se sont avérées trop importantes. Dans la plupart des cas l'état stationnaire ne pouvait pas être atteint sur la longueur du bassin. D'autre part la gîte obtenue dépend d'une supposition sur la hauteur du centre vélique et du centre de gravité. Les résultats ne peuvent pas être étendus à des cas de chargement différents. Cette technique ne pourra donc pas s'inscrire dans une démarche d'optimisation ou de détermination de l'influence de paramètres de réglage. Elle a été abandonnée.

C. Talotte au cours de sa thèse [23] a ensuite utilisé un dispositif composé d'un dynamomètre à six composantes sur lequel la maquette est fixée par un orienteur. Le pilonnement et l'assiette sont libres. Le point de poussée est abaissé au niveau du pont et le problème est donc de compenser le déficit de moment piqueur pour que l'assiette en essai soit proche de l'assiette en fonctionnement réel. La procédure impose donc une série d'essais préalables en traction droite à assiette fixe et à différente vitesse pour obtenir une courbe de traînée qui permet de calculer une compensation de moment. Ce calcul dépend du modèle vélique pour évaluer l'altitude du centre d'efforts. La compensation est ensuite obtenue en déplaçant longitudinalement une masse sur le pont. Cette méthode semi-captive est intéressante lorsqu'il s'agit de comparer rapidement et au moindre coût des carènes entre elles à des points de fonctionnement donnés. Elle ne permet pas de déterminer l'influence de l'assiette sur la performance alors que nous avons montré l'importance de ce paramètre [21]. En définitive, après avoir réalisé quelques essais avec cette méthode nous avons préféré nous orienter vers des méthodes, qui conformément à notre démarche, permettent de caractériser la carène indépendamment de tout modèle d'effort vélique.

### *4.5 LES METHODES EXPERIMENTALES UTILISEES*

#### 4.5.1 La méthode « modèle fixe »

Cinq paramètres influencent le torseur des efforts hydrodynamiques sur une carène en traction rectiligne et uniforme : la vitesse d'avance, l'enfoncement par rapport à la surface libre, et trois angles caractérisant l'attitude du modèle : la gîte, l'assiette et la dérive.
Dans un premier temps nous avons fait varier ces cinq paramètres indépendamment afin de mesurer l'influence de chacun d'entre eux. Pour chaque essai le modèle est lié rigidement au système de mesure. Avec cette technique on a donc cinq paramètres de réglage et six mesures pour la détermination complète du torseur des efforts sur la carène dans chaque configuration. Cela conduit à un nombre très élevé d'essais mais également à des difficultés de mise en œuvre. L'enfoncement notamment est délicat à régler, car la composante verticale des efforts sur la maquette diffère grandement selon la valeur des autres paramètres de réglage. Pour nombre d'essais le déplacement effectif du modèle s'avère irréaliste. Toutes ces raisons nous ont fait abandonner cette méthode.

#### 4.5.2 La méthode semi captive avec pilonnement libre

Seul le degré de liberté de translation verticale est laissé au modèle. L'assiette est fixe, ce qui d'une part permet de caractériser son influence et d'autre part donne des mesures plus stables. Nous avons adopté un point de traction fixe au niveau du centre de gravité de la carène. Les déformations engendrées par les moments de torsion sont évaluées et on en tient compte dans le dépouillement. Il reste cinq paramètres de réglage. On règle directement le déplacement du modèle et non plus l'enfoncement. Ce qui permet d'obtenir des résultats pour des configuration plus réalistes.

### *4.6 LA METHODE DES PLANS D'EXPERIENCES*

Notre objectif est de modéliser les grandeurs mises en jeu (le torseur des efforts appliqués à la carène) à partir de la connaissance d'un minimum de points de fonctionnements. Les efforts hydrodynamiques sur la carène en gîte et dérive à vitesse constante sont obtenus à partir de différentes méthodes d'interpolation entre des points de fonctionnements issus des campagnes d'essais en bassin des carènes.

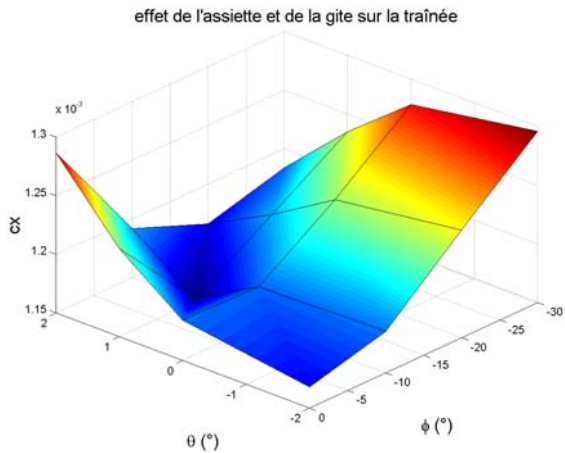
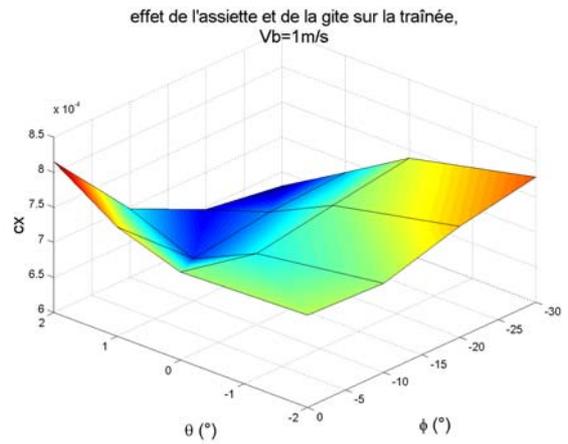

**Figure 2**     **Figure 3**

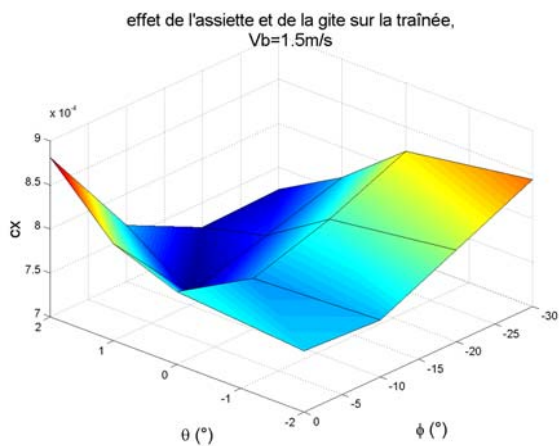
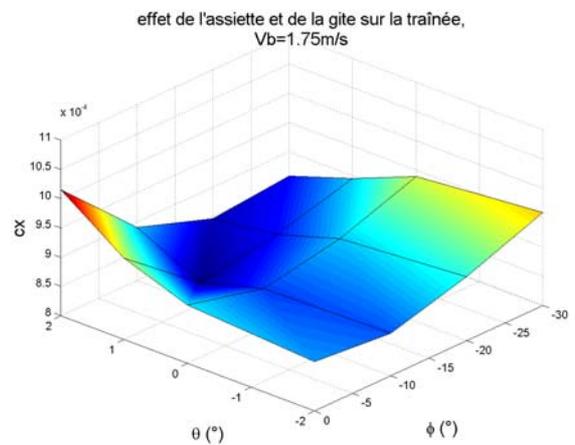

**Figure 4**     **Figure 5**

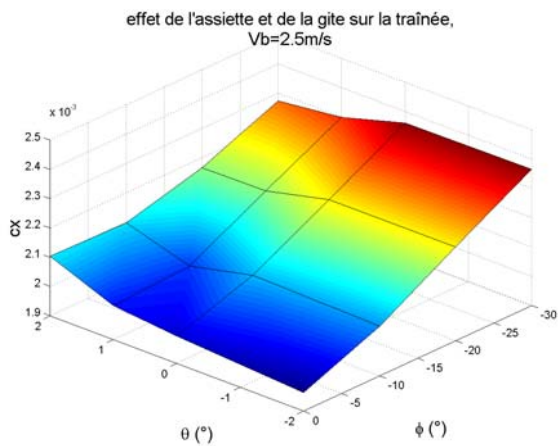
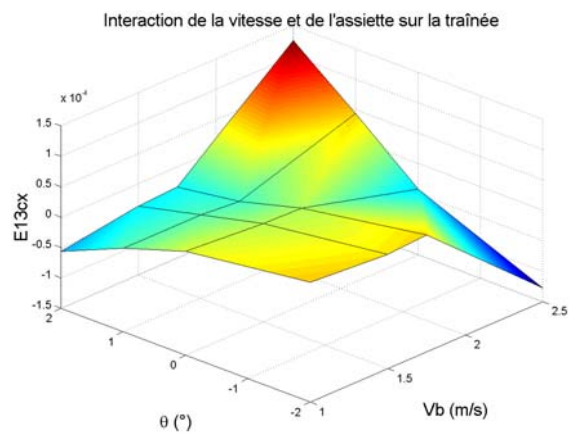

**Figure 6**     **Figure 7**

Les outils d'interpolations courants (linéaire, splines, splines cubiques), permettent de résoudre ce problème à l'aide d'un maillage complet de l'espace vectoriel à n dimensions, où n est le nombre de facteurs qui influencent la grandeur étudiée. Pour nos modélisations le nombre de facteurs est tel, que réaliser un maillage complet du domaine étudié se révèle trop coûteux en temps de calcul ou d'expérimentation. La méthode que nous avons employée pour choisir ces points de fonctionnements est celle des plans d'expériences. La théorie des plans d'expériences, très utilisée aujourd'hui dans les milieux techniques et industriels, fournit outre une approche structurée, des outils d'optimisation du ratio informations recueillies sur le nombre d'expériences. Typiquement, notre objectif est de limiter le nombre de configurations d'essais à 128 par carène pour obtenir les efforts hydrodynamiques à vitesse constante dans toutes les attitudes possibles rencontrées en navigation. Cela représente approximativement 64 heures d'essais.

Pour passer des résultats sur le modèle au réel, on suppose que la composante de pression des efforts est proportionnelle à $\lambda^3$, avec $\lambda$ le rapport d'échelle de longueur entre la maquette et le réel. On utilise la méthode préconisée par l'International Towing Tank Conference pour l'évaluation de la composante de friction et la méthode de Prohaska pour la détermination du coefficient de forme.

### 4.6.1 Décomposition

Les efforts hydrodynamiques sont considérés comme des fonctions à cinq variables. Chaque fonction est décomposée comme la somme de plusieurs fonctions à une, deux, trois, quatre et cinq variables.

$$\begin{aligned}
f_x(Vb,\phi,\theta,\beta,\Delta) = & \overline{f_x} + E1 f_x(Vb) + E2 f_x(\phi) + E3 f_x(\theta) + E4 f_x(\beta) + E5 f_x(\Delta) & \}\textit{Effets simples} \\
& + E12 f_x(Vb,\phi) + E13 f_x(Vb,\theta) + E14 f_x(Vb,\beta) + \cdots & \}\textit{Interactions d'ordre 2} \\
& + E123 f_x(Vb,\phi,\theta) + E124 f_x(Vb,\phi,\beta) + \cdots & \}\textit{Interactions d'ordre 3} \\
& + E1234 f_x(Vb,\phi,\theta,\beta) + E1235 f_x(Vb,\phi,\theta,\Delta) + \cdots & \}\textit{Interactions d'ordre 4} \\
& + E12345 f_x(Vb,\phi,\theta,\beta) & \}\textit{Interactions d'ordre 5}
\end{aligned}$$

La méthode ne présente d'intérêt que si les termes d'ordre supérieur sont négligeables. On se limite donc au calcul des effets simples et aux interactions d'ordre 2 voire 3. De plus, les coefficients sont obtenus par des calculs de moyenne ce qui améliore la partie aléatoire de la précision des mesures.

Nous avons implémenté dans le simulateur le modèle Tagushi de premier ordre pour ce qui concerne la traînée de la carène et de la quille. Pour la portance un modèle de premier ordre ne suffit pas car il pose des problèmes de continuité à gîte et dérive nulle. En effet, nous n'avons exploré dans nos campagnes d'essais qu'un quart de l'espace vectoriel des 5 paramètres retenus en considérant que les efforts sont symétriques par rapport à la gîte et par rapport à la dérive. Cette hypothèse ne devrait pas être trop pénalisante car le voilier ne parcourt pas souvent les zones du domaine que nous n'avons pas exploré (gîte et dérive de signe opposé). Toujours est-il que pour les fonctions impaires (la portance et certains moments) le modèle additif simple ne peut garantir une continuité à zéro. En attendant d'implémenter des modèles d'ordre supérieur nous avons choisi de conserver pour la portance le modèle issu de la théorie des ailes proposé par P. Van Oossanen [24]. De plus, le couplage entre les vitesses de rotation de roulis et de lacet sont plus facilement pris en compte en distinguant les efforts sur la carène et les efforts sur la quille.

### 4.6.2 Analyse des résultats

On peut utiliser une troncature du modèle pour caractériser l'influence d'un paramètre ou d'une combinaison de paramètres sur un effort. La Figure 2 représente par exemple la somme $S = \overline{C_x} + E2 c_x + E3 c_x + E23 c_x$, on isole ainsi les effets de la gîte et de l'assiette sur la traînée.

On peut noter que le point optimal, pour lequel le coefficient de traînée est minimal, ne se situe pas carène droite mais avec un peu de gîte (entre 10 et 15°) et une assiette légèrement piqueuse (environ 1°). Il y a une évolution combinée de l'influence des 2 paramètres, l'assiette optimale dépend de la gîte et inversement.

Cette figure est une coupe de l'évolution du Cx au point moyen $M(Vb_M, \phi_M, \theta_M, \beta_M, \Delta_M)$, de l'espace vectoriel d'investigation. en ce point : $Cx(Vb_M, \phi_M, \theta_M, \beta_M, \Delta_M) = \overline{Cx}$

La coupe a une allure différente selon le lieu. Les coupes du coefficient de traînée sont ici représentés pour 4 vitesses (1, 1.5,1.75 et 2.5 m/s). A faible vitesse(1m/s) la gîte combinée à une assiette piqueuse (maquette enfoncée sur l'avant) améliorent notablement le coefficient de traînée, tandis que la tendance s'inverse à haute vitesse (2.5m/s). L' interaction entre la vitesse et l'assiette est représentée dans le modèle par la fonction "E13cx" (Figure 7).

### 4.6.3 Erreur de modélisation sur le coefficient de traînée Cx

|  | Réponse par effet simple | Réponse avec intéraction ordre2 interpolation linéaire | Réponse avec intéraction ordre3 interpolation linéaire |
|---|---|---|---|
|  | erreur relative | erreur relative | erreur relative |
| erreur moyenne | 4.7% | 2.3% | 1.9% |
| ecarttype | 6.3% | 3.5% | 2.6% |

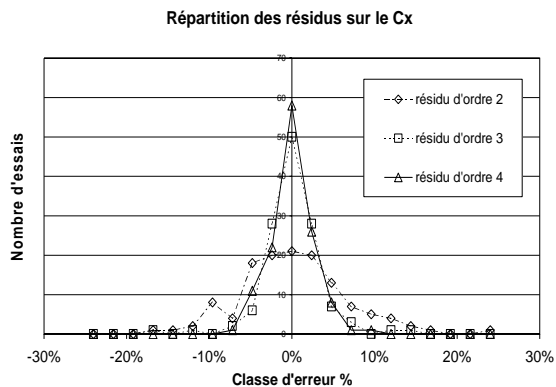

**Figure 8**

La différence entre le modèle d'ordre n et les résultats expérimentaux est appelée résidu d'ordre n. Plus l'ordre du modèle est élevé, plus le résidu (l'erreur) est distribué suivant une courbe de Gauss. On considère couramment que les intéractions d'ordre élevée (>3) sont faibles et se confondent pour un plan factoriel avec la part aléatoire de l'erreur de mesure.
On observe que l'erreur absolue (en Newton) est comparable avec la précision du système de mesure pour des essais à moyenne vitesse.

## 5  EFFORTS AERODYNAMIQUES

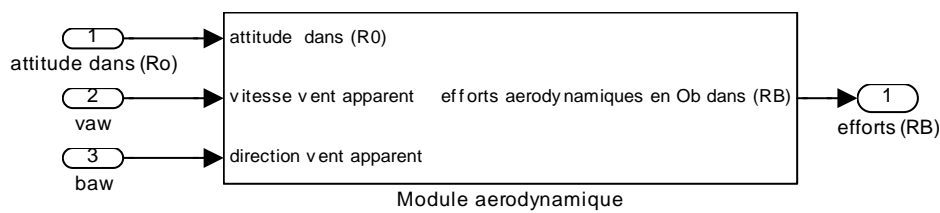

**Figure 9**

La modélisation des efforts aérodynamiques a longtemps été le parent pauvre des études sur les voiliers. En plus des problèmes théoriques et expérimentaux qu'elle pose, cette modélisation présente des difficultés du fait des multiples réglages possibles de la structure souple que constituent les voiles et le gréement. Il faut donc veiller à ce que l'évaluation des efforts véliques ne soit pas "précisément fausse". Il faut également remarquer que dans la démarche de conception, une évaluation sommaire de ces efforts, si elle est faite de façon cohérente, peut suffire à comparer des performances brutes de voiliers. Récemment les études numériques se sont développées, d'abord en fluides parfaits [17] [3] [4] [9] [12] [13]et plus récemment en fluide visqueux [4].Si les méthodes "Fluide Parfait" évaluent correctement la portance au près serré, lorsque les voiles travaillent en finesse, il n'en va pas de même de la traînée qui est largement affectée par les décollements tourbillonnaires à la jonction du mat et de la grand voile. Si bien que les codes fluides parfaits ne peuvent être exploités pour évaluer la performance sans être associés à des résultats expérimentaux. On a donc cantonné ces méthodes soit à l'étude des déformations [4], soit à l'étude de l'interaction entre les voiles [3] [17]. A l'aide des nouveaux outils informatiques, on commence à envisager l'étude en fluide visqueux. Mais les temps de calcul sont importants. Caponnetto et al. [4] donnent un ordre de grandeur de 8 heures pour la convergence du calcul sur les 12 processeurs d'une station Silicon Graphic origin2000. Pour les premières versions du simulateur nous avons opté pour une solution simple et éprouvée. Nous avons adapté le modèle empirique de Myers datant de 1975 [19], et repris dans de nombreux VPP [24].

La force latérale au près est définie par :

$$F_L = \frac{\rho_A \cdot V_{wA}^2}{2} \cdot A_S \cdot \frac{C_L}{\cos\gamma} \cdot \cos(\beta - \gamma) \cdot \cos\phi$$

Myers donne une valeur indicative de 10° pour γ. Van Oossanen donne une évaluation systématique de $\gamma = 0.044 + \dfrac{0.089 \cdot C_L^2}{1 - \phi}$

L'expression des coefficients de portance est déterminée par des formules de régression polynomiale au second ordre en fonction de l'angle d'incidence.

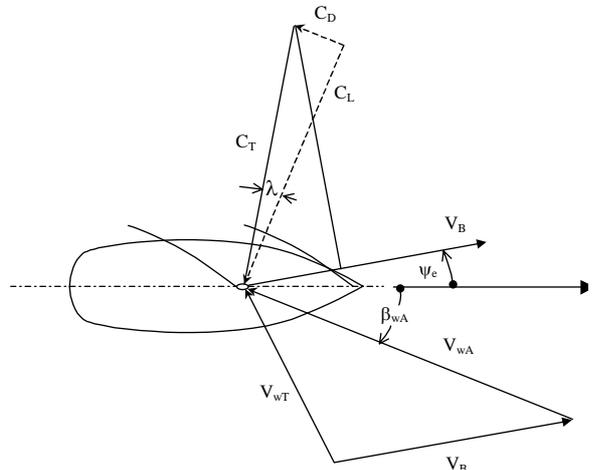

Pour le génois, par exemple le modèle donne :

$$C_L = 1.04 + 0.344 \cdot \beta_{WA} + 0.041 \cdot \beta_{WA}^2; \quad avec \quad \beta_{WA} \in \left[\frac{\pi}{8}; \frac{\pi}{2}\right] \text{ (génois)}$$

Ce modèle ne fait pas intervenir de paramètres de réglages. Il est donc intéressant pour les premières versions d'un simulateur. En effet, l'intérêt d'introduire des réglages de voiles sophistiqués est discutable tant que les modèles utilisés dans les autres modules ne seront pas suffisamment fins pour que l'influence de ces réglages soit significative. Dans la littérature [10][20][22], le réglage des voiles est introduit sous forme de deux coefficients multiplicatifs, l'un pour réduire la voilure, il affecte donc la position du centre de poussée vélique et la surface de la voilure, tandis que le deuxième affecte les coefficients aérodynamiques pour tenir compte d'un réglage du creux des voiles. Ils pourront donc être introduits facilement dans le modèle par la suite.

Dans un premier temps nous avons adopter le modèle simple de Myers [19]. Nous avons souhaité ensuite confronter des éléments du modèle au calcul numérique en utilisant le code développé par l'équipe de M. Guilbaud [9][17].

### 5.1 LE MODELE AERODYNAMIQUE DANS UN SIMULATEUR

Un simulateur est le plus souvent piloté par une seule personne et il n'est pas utile de reproduire toute la complexité du réglage du voilier dont plusieurs " régleurs "s'occupent en navigation réelle. Le modèle aérodynamique dans un simulateur doit donc être simple d'utilisation si on souhaite reproduire le même environnement dans le simulateur et libérer le " pilote " des contraintes du réglage. La grande majorité des modèles de la littérature adopte cette optique. On dénombre au plus trois paramètres de réglage : le vrillage [7] [16], la réduction de voilure, le creux. L'influence de ces paramètres doit être évaluée au préalable pour qu'ils puissent intervenir globalement dans le simulateur comme de simples gains agissant sur les efforts véliques et leur point d'application.

### 5.2 COMPARAISON DU MODELE DE MYERS AVEC UN CODE NUMERIQUE "FLUIDE PARFAIT"

Tout régatier connaît l'importance des effets visqueux et des décollements de l'écoulement autour d'une voile. L'utilisation d'un code fluide parfait ne se justifie qu'à l'allure du près où les voiles travaillent en finesse et pour la détermination de la portance. Caponneto [3] confirme qu'à cette allure la méthode fluide parfait donne de bons résultats pour le coefficient de portance.

Comme dans ces conditions, la plupart des modèles véliques considèrent simplement que l'effort vélique est perpendiculaire au mat, et la position du centre de poussée est évaluée par des considérations géométriques sommaires, nous avons voulu vérifier si un calcul fluide parfait donnait de meilleures indications d'une part sur le centre de poussée et l'influence éventuelle des différents paramètres de réglage (incidence, vrillage), et d'autre part sur la direction de l'effort vélique.

Les premiers résultats ont montré que l'utilisation de la méthode fluide parfait impliquait des contraintes encore plus restrictives que nous ne l'avions escompté. La première condition pour qu'un calcul " fluide parfait " soit exploitable est que la voile doit être adaptée partout. Cela signifie que l'on doit vérifier qu'il n'y a pas de contournement des arêtes vives (guindant, bordure, têtière). S'il existe dans le calcul un point de contournement, le coefficient de pression locale prends des valeurs qui ne sont pas réalistes et le résultat est alors très sensible à la finesse du maillage. Dans ce cas il n'y a pas convergence entre le calcul d'efforts par intégration des pressions et par circulation. Il est dès lors apparu impossible d'adapter sans en modifier la géométrie les profils de voiles fournis par le fabricant. Il a donc été nécessaire de modifier la géométrie des voiles en modifiant le vrillage.

L'angle de vrillage est un angle de rotation de la voile autour de son guindant qui varie linéairement en fonction de l'altitude y. Il est nul au niveau le plus bas de la voile ($y_{min}$), dans le repère ayant son guindant pour axe vertical, et il est maximal au point le plus haut ($y_{max}$).

$$vrillage(y) = \frac{(y - y_{\min})}{(y_{\max} - y_{\min})} \cdot vrillage_{\max}$$

Nous avons comparé nos résultats avec ceux du modèle de Myers [19] et avec les formules 2D de la jauge IMS[5] [7]. Nos calculs se situent logiquement en dessous de la jauge IMS 2D puisque notre gréement a une envergure finie.

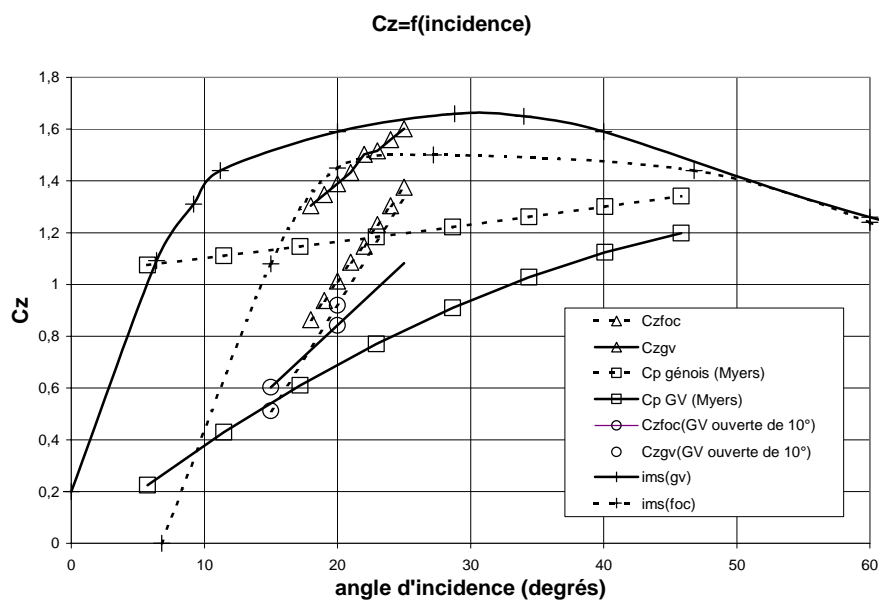

**Figure 10**

Le modèle de Myers est un modèle auto-réglé, c'est à dire qu'il considère un réglage optimal de la voilure pour chaque angle d'incidence. Notre calcul s'effectue sur une géométrie fixe de la voilure que nous faisons simplement varier en incidence par rapport à l'écoulement infini amont. Les courbes n'ont naturellement pas la même allure. Elles se coupent simplement au point de fonctionnement du voilier au près. Pour le foc l'intersection se situe aux alentours de 23° d'incidence par rapport au vent apparent. Pour la grand-voile il existe une différence très sensible selon que l'on borde la grand voile dans l'axe ou si on l'ouvre de seulement 10°. Il est donc délicat d'effectuer une comparaison. On note d'ailleurs que le modèle de Myers et celui de l'IMS se contredisent puisque l'un place le coefficient de pression de la grand voile nettement en dessous de celui du foc alors que l'autre le place au-dessus. On peut simplement conclure que nos résultats semblent raisonnables.

Ces premiers résultats nous ont montré que la méthode fluide parfait a un domaine d'application restreint. On ne peut pas se contenter de calculer un écoulement autour d'une géométrie donnée sans vérifier que le calcul est valide vis à vis des hypothèses et notamment vis à vis du non contournement des arêtes vives. Le cas échéant, il faut modifier cette géométrie. Nous plaçons en perspective l'utilisation d'un algorithme systématique de modification de la géométrie des voiles (vrillage et incidence) afin d'obtenir des résultats pour des profils adaptés à l'écoulement. Nous pourrions alors déterminer la plage exacte de validité de la méthode en comparant de nouveau nos résultats aux modèles auto-réglés de Myers et de l'IMS.

### *5.3 INTERACTION AERODYNAMIQUE*

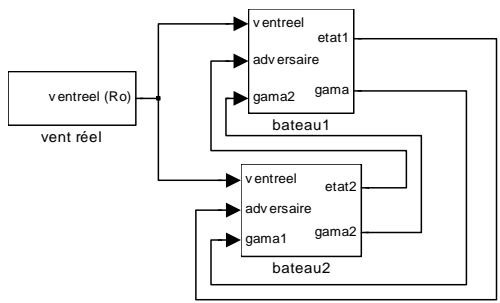

**Figure 11**

Cet aspect constitue un des principaux points faibles des simulateurs présents dans le commerce, il est cependant une composante essentielle de la tactique dans une régate au contact.

La mise en œuvre du code fluide parfait s'est révélée plus compliquée que prévue du fait des problèmes évoqués plus haut et nous souhaitons valider la procédure d'utilisation d'un code fluide parfait avant d'évaluer l'interaction entre deux voiliers avec cette méthode. Nous avons introduit pour le moment, un modèle très schématique pour l'interaction aérodynamique. On se ramène à un problème bidimensionnel dans le plan horizontal ($\vec{X_0}, \vec{Y_0}$). La présence de l'adversaire est représentée par un tourbillon dont on obtient l'intensité à partir de la portance développée par ses voiles :

$$\Gamma = -\frac{\text{portance}}{h \cdot \rho_A \cdot V_{WA}}$$ avec *h qui représente l'envergure du gréement.*

Le champ de vitesse de perturbation est le champ induit par ce tourbillon. Il est additionné au champ de vitesse incident pour obtenir la vitesse.

$$\vec{V_{WT}} = \vec{V_{WT\infty}} + k \frac{\Gamma}{2 \cdot \pi \cdot r^2} \begin{vmatrix} -(y - y_{adversaire}) \\ (x - x_{adversaire}) \end{vmatrix}$$

Le coefficient k permettra d'ajuster le modèle grâce aux résultats disponibles dans la littérature et en particulier ceux de Caponnetto [3].

## 6 SIMULATION

Nous présentons ici un cas simple de simulation (Figure 12). Deux voiliers louvoyant sont placés dans un champ de vent constant de 5 nœuds orienté à -90 degrés par rapport à l'axe. Chacun d'entre eux est piloté par un correcteur proportionnel asservi par rapport à un angle du vent apparent correspondant approximativement au VMG (vitesse optimale de remontée au près). A l'instant initial les voiliers partent arrêtés. Le voilier de gauche part avec un gain au vent de 4 mètres soit une demi-longueur et termine avec plus de 10 mètres d'avance après 250 secondes.

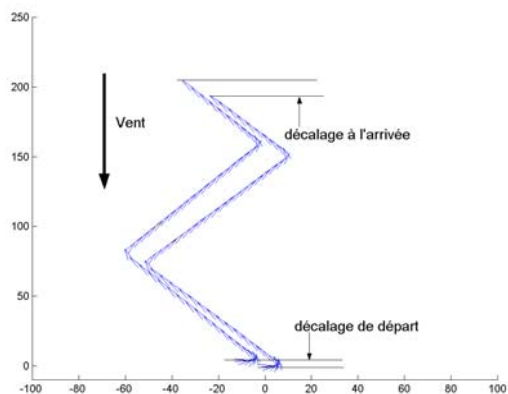 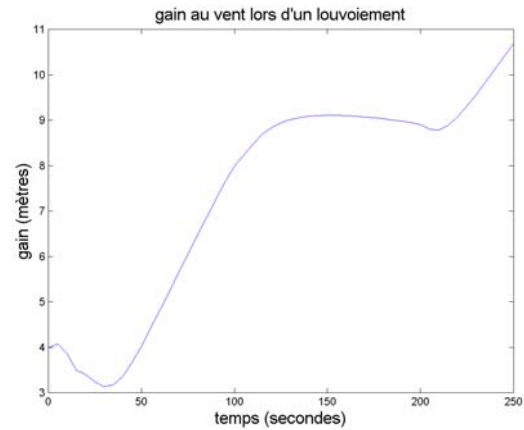

**Figure 12**                           **Figure 13**

La Figure 13 montre que le voilier de tête accentue d'avantage son avance lorsqu'il se trouve sous le vent de son adversaire et qu'il a plutôt tendance à perdre du terrain ou à ne plus en gagner lorsqu'il se trouve au vent. Ce phénomène est bien connu des régatiers sous le nom de "la position favorable sous le vent". Il est ici accentué par le fait que le sillage visqueux qui défavoriserait d'avantage encore le voilier en arrière n'est pas modélisé.

## 7    CONCLUSION

Une nouvelle démarche a été mise en œuvre pour l'exploitation des campagnes d'essais en bassin sur un voilier. Cette démarche nous permet d'avoir une représentation exhaustive de l'influence des paramètres de vitesse de translation , de position et d'attitude stationnaires sur les efforts hydrodynamiques. Elle fournit en outre une modélisation simple et rapide compatible avec la simulation en temps réel.

Un simulateur a été construit qui met en interaction deux voiliers. Les premiers résultats sont en accord avec les observations du régatier, même s'il reste à introduire quelques améliorations pour une représentation fidèle du comportement du voilier.
Dans le cours terme sont prévus :
- l'introduction des interactions d'ordre 2 pour le modèle hydrodynamique,
- l'introduction d'une modélisation non linéaire de l'hydrostatique, qui s'impose si nous voulons pleinement exploiter les résultats des campagnes d'essais,
- l'introduction de la modélisation d'un sillage visqueux pour l'interaction aérodynamique et ensuite l'exploitation des résultats d'un code numérique tridimensionnel.

A plus long terme, notamment grâce au progrès des calculs numériques tridimensionnels en fluide réel, on peut envisager d'évaluer plus finement l'influence des vitesses de rotation du voilier sur la résistance à l'avancement et sur la manœuvrabilité.